\documentclass[preprintnumbers,secnumarabic,amsmath,amssymb,nofootinbib,floatfix]{revtex4}
\usepackage{varioref,exscale,latexsym,amsmath,amssymb}
\usepackage{graphicx}

\usepackage{graphicx}
\usepackage{slashed}
\usepackage{dcolumn}
\usepackage{bm}

\def\beq{\begin{equation}}

\def\eeq{\end{equation}}

\def\beqa{\begin{eqnarray}}

\def\eeqa{\end{eqnarray}}

\begin{document}
\preprint{ACFI-T17-04}

\title{{\bf Quartic propagators, negative norms and the physical spectrum }}

\medskip\

\author{ John F. Donoghue${}$}
\email[Email: ]{donoghue@physics.umass.edu}
\affiliation{~\\
Department of Physics,
University of Massachusetts\\
Amherst, MA  01003, USA\\
 }

\begin{abstract}
Many arguments against quartic propagators, negative norm states and related effects concern the sicknesses which occur when the spectrum of
the free particle Hamiltonian is formed. However, if the theory is more complicated, for example involving confinement such that the particle in question does not appear in the physical spectrum,
those considerations do not apply directly. Path integral methods suggest that some of these may be acceptable theories. I provide an example that
should be able to be simulated on a lattice which then allows a non-perturbative resolution of this question. In its SU(2) version it involves a scalar triplet with a quartic derivative Lagrangian coupled to the SU(2) gauge field. If this is verified to be a healthy theory, it could open new avenues in model building. I also discuss how strong interactions can dynamically modify the dispersion relation leaving a healthy effective field theory, using conformal gravity coupled to a Yang-Mills theory as an example. Such a theory could possibly form a UV completion for quantum gravity.
\end{abstract}
\maketitle

\section{Pathologies and the physical spectrum}

The standard analysis of the content of a field theory starts with the free field theory. From the free Lagrangian one constructs the Hamiltonian and finds the physical energy spectrum of the theory. When interactions are introduced at a later stage, one studies how they renormalize the mass of the original particles, but the dispersion relations and Hilbert space are set by the original free Hamiltonian. Theories which prove to be sick at this stage are discarded.

However, we are also now familiar with many theories whose free Lagrangian has no relation to the physical spectrum. An important class involves confining theories. Here the original particles of the Lagrangian do not even appear in the spectrum. In other cases, interactions can generate emergent interactions or degrees of freedom, which can form an effective field theory below the scale of the original theory. Study of the properties of the free field theory are of little value in these cases. Here we propose that some theories whose free field construction is sick may nevetherless be valid quantum field theories if the sick features are not manifest in the physical spectrum.

Among the theories which are discarded by the free field analysis are those with higher time derivatives in the Lagrangian. Ostrogradsky's theorem\cite{Woodard:2015zca} says that such theories always have instabilities when the Hamiltonian is constructed. Because of the extra time derivatives, there are extra canonical momenta. For example, for a theory with two time derivatives acting on a field, i.e. ${\cal L} = {\cal L}(\phi, \dot{\phi}, {\ddot{\phi}}) $, arranged in the form
\begin{equation}
{\cal L} = -\frac12 \Box \phi \Box \phi +...
\end{equation}
there are two coordinates $\phi$ and $\dot{\phi} $ and two canonical momenta,
\begin{eqnarray}
\pi_1 &=& \frac{\partial {\cal L}}{\partial \dot{\phi}} -\frac{d}{dt}\frac{\partial {\cal L}}{\partial \ddot{\phi}}   \nonumber  \\
\pi_2 &=& \frac{\partial {\cal L}}{\partial \ddot{\phi}} \ \ .
\end{eqnarray}
Ostrogradsky's construction leads to a Hamiltonian
\begin{equation}
{\cal H} = \pi_1 \dot{\phi} + \pi_2 a(\phi, \dot{\phi}, \pi_2) - {\cal L} (\phi, \dot{\phi}, a(\phi, \dot{\phi}, \pi_2) )
\end{equation}
where $a(\phi, \dot{\phi}, \pi_2)$ is an acceleration which does not depend on $\pi_1$. Because $\dot{\phi}$ is an independent coordinate, one
cannot remove $\dot{\phi}$ in the first term to write this as $\pi_1^2$. The fact that the Hamiltionian is linear in $\pi_1$ and that there are no other compensating powers of $\pi_1$ elsewhere in the Hamiltionian, indicates an instability in defining the energy and the time dependence. The presence of interactions in the action would presumably excite this unstable mode.

Moreover in Lorentz invariant theories that contain higher derivatives the propagators are no longer quadratic. This leads to propagators which go asymptotically as $1/q^4$.  An analysis of the quartic propagator is often considered in the presence of a quadratic term also,
\begin{equation}\label{quartic}
\frac{-i}{q^4} \sim \frac{-i}{q^2(q^2-\mu^2)} = \frac{1}{\mu^2}\left(\frac{i}{q^2}- \frac{i}{q^2-\mu^2}\right)
\end{equation}
which can be reduced to a pair of quadratic propagators, of which one has the wrong overall sign, indicating a negative norm state - a ghost. These features lead most researchers\footnote{Some exceptions to the general consensus include \cite{Tomboulis:1983sw, Smilga, Bender:2007wu, Mottola:2016mpl, Holdom, Salvio:2015gsi, Donoghue:2016vck, Donoghue:2016xnh}} to discard quartic derivative Lagrangians as being sick.

However these sicknesses does not seem to be shared by the path integral in all cases. For example, consider a theory with a
Lagrangian
\begin{equation}
{\cal L}= -\frac12 (D^2\phi)^\dagger D^2 \phi +...
\end{equation}
where we here can include some gauge interaction with a covariant derivative $D_\mu$. This would be a theory that should exhibit the
Ostrogradsky instability. In a path integral treatment, on can perform the
quadratic path integral over $\phi$ obtaining the result
\begin{equation}
\frac{1}{[{\rm det}(D^2D^2)]^{1/2}} = e^{-\frac12 Tr \log (D^2D^2)} = e^{-\frac12\int d^4x <x|Tr \log (D^2D^2)|x>}
\end{equation}
However, using the property of the logarithm, one has
\begin{equation}
\log (D^2D^2) = 2 \log (D^2)
\end{equation}
so that the resulting action at this level is just twice the action that would result from a quadratic action. The Hamiltonian instability does not appear in the path integral.

Moreover, a theory with a quartic propagator can make an acceptable perturbation theory in the ultraviolet, again using path integrals rather than
the free Hamiltonian to set up the theory. If one considers the generating functional
\begin{equation}
Z_0[J] = \int [d\phi] \exp i\int d^4x~\left[- \frac12 \Box \phi \Box \phi - J\phi \right]
\end{equation}
it can be solved by completing the square in the usual fashion, such that
\begin{equation}
Z_0[J] = Z_0[0] \exp i\int d^4x d^4y ~ \frac12 J(x) \Delta(x-y) J(y)
\end{equation}
where
\begin{equation}
\Delta(x-y) = <x| \frac{1}{\Box^2}|y> = \int d^4q~ \frac{e^{-iq\cdot (x-y)}}{q^4}
\end{equation}
In the presence of interactions, the generating functional can lead to the perturbative expansion in the usual way, without obvious signs of
being a sick theory. Of course, there may be enhanced IR divergences in such a theory, because of the propagator growth at low momentum. But the severity of these depends on how soft the interactions are in the IR, and are theory specific. In addition, confinement may resolve the IR issues by not allowing free propagation in the IR.

We see here that the Hamiltonian approach to a theory and the path integral approach appear to indicate different fates for theories with higher derivatives. While the equivalence of the two approaches is known for simple field theories with the usual derivative structure, to our knowledge it has not been proven for theories with higher derivatives.

These considerations indicate the likelihood that a theory containing a field with a quartic propagator could be a valid QFT if the physical spectrum of the theory is acceptable, either because of confinement or because interactions change the character of the dispersion relations for the physical particles.

\section{An example with confinement}

One direct way to evade Ostrogradsky's theorem is to have a theory where the fundamental particles are confined. The free Hamiltonian formed from the original theory is meaningless as a statement of the physical spectrum. Of course, this feature is not a proof that such theories {\em are} good quantum field theories, although the path integral arguments suggest that some of them may be. But it does mean that present discussion of such theories are inadequate.

Because confining theories are not under analytic control, it would be useful to have an example which can be simulated on a lattice.
Here is one such construction. Consider a scalar field which carries a Yang-Mills charge. For definiteness, let us consider the gauge group to be
SU(N) and the scalar, $\phi^a, ~~a=(1,2,...,N)$, to be a adjoint representation, arranged in an SU(N) valued matrix
\begin{equation}
U = e^{it^a \frac{\phi^a}{f}}
\end{equation}
where $t^a$ are the $N\times N $ traceless Hermetian matrices related to the group generators. This has the gauge transformation
\begin{equation}
U \to V(x)UV^\dagger(x)
\end{equation}
with $V(x)$ in SU(N).
If we ask for a scale-invariant action which is gauge invariant, we arrive at
\begin{eqnarray}\label{quarticL}
{\cal L} &=& -\frac{1}{4g^2} F^a_{\mu\nu}F^{a\mu\nu} -\frac{f^2}{4} Tr\left[(D^\mu D_\mu U)^\dagger (D^\nu D_\nu U) \right]  \nonumber  \\
 &+& d_1\left(Tr\left[D^\mu U^\dagger D_\mu U  \right]\right)^2 +d_2Tr\left[D^\mu U^\dagger D^\nu U  \right]Tr\left[D_\mu U^\dagger D_\mu U  \right] \ \  \nonumber \\
 &+& d_3 Tr\left[D^\mu U^\dagger D^\nu U  D_\mu U^\dagger D_\mu U \right] + d_4 Tr\left[U^\dagger D^2 U  \right]Tr\left[D_\mu U^\dagger D^\mu U \right]  \ \  \nonumber \\
 &+& d_5 Tr\left[U^\dagger D^2 U  D_\mu U^\dagger D^\mu U \right] +  d_6 Tr\left[U^\dagger D^2 U  \right] Tr\left[U^\dagger D^2 U  \right] \ \  \nonumber \\
 &+& d_7 F^a_{\mu\nu} \left(Tr\left[\tau^aD^\mu U^\dagger D^\nu U  \right]+ Tr\left[D^\mu U^\dagger\tau^a D^\nu U  \right]\right) \ \nonumber \\
  &+& d_8 F^a_{\mu\nu}F^{b\mu\nu} \left(Tr \left[\tau^a U^\dagger\tau^b U\right]- 2\delta^{ab}\right)
\end{eqnarray}
where the covariant derivative is defined via
\begin{equation}\label{covD}
  D_\mu  U =\partial_\mu U +\frac{i}{2} [t^aA^a_\mu , U]   \ \ .
\end{equation}
Connoisseurs of non-linear actions will recognize that the action of Eq. \ref{quarticL} is related to the SU(N) chiral Lagrangian originally written down by Gasser and Leutwyler \cite{Gasser}. The chiral Lagrangian with two derivatives has been eliminated by the imposition of scale invariance. The Gasser-Leutwyler construction used the lowest-order equation of motion in constructing the Lagrangian, which is not appropriate in the present case. This accounts for the factors of $D^2U $ seen in the present Lagrangian. For SU(2), the term proportional to $d_3$ can be eliminated by an SU(2) identity. Finally there are no explicit mass terms considered here - the scalars do not have a mass. Like the Gasser-Leutwyler construction Eq. \ref{quarticL} also has a {\em global} $SU(N)_L \times SU(N)_R$ invariance $U\to LUR^\dagger$ which constrains the action\footnote{In constructing the global invariance, one temporarily endows the gague field with global left and right handed properties \cite{Gasser}.}.

Because of the four derivatives in the kinetic energy terms, this theory would have the Ostrogradsky instability in the free field limit. However,
the scalar field will be confined and will not appear in the spectrum. In a lattice simulation, one would test whether the correlation functions are well behaved as in a standard confined theory, or if they indicate long range instabilities.

The one-loop renormalization involving the scalar field can be performed and does not display a pathology. For the gauge interaction, one can expand the action to quadratic order in the scalar field, obtaining
\begin{equation}
{\cal L}= -\frac12 \phi^a D^2 D^2 \phi^a +...
\end{equation}
Using the reasoning described above, the result is just twice that of a normal scalar field.
In particular, the one loop-divergence is
\begin{equation}
S_{div} = \int d^4x ~\frac{1}{\epsilon}\frac{2a_2}{16\pi^2} = \int d^4x~ \frac{C_2}{48\pi^2\epsilon} F^a_{\mu \nu}F^{a\mu\nu}
\end{equation}
where $a_2$ is the second Seeley-DeWitt coefficient and $C_2$ is the quadratic Casimir for the scalar field (with $C_2 =2$ for the tripet representation in SU(2)). The beta function within SU(2) is then
\begin{equation}
\beta(g) = -\left[\frac{11N}{3} - \frac{1}{4}C_2\right]\frac{g^3}{16\pi^2}  \ \ .
\end{equation}
The theory is asymptotically free in the gauge coupling.

Despite being non-linear in the scalar field, the theory is renomalizeable in the scalar sector also. Power-counting proceeds differently in a theory with quartic propagators. The result can be seen most simply by noting that the Lagrangian is scale invariant. When regularized dimesionally, the UV divergences will then be local objects of dimension four, implying that they have the structure of the most general gauge invariant Lagrangian already given in Eq. \ref{quarticL}. Wavefunction or coupling constant renormalization will then handle the divergences. In more detail, the renormalization due to scalar loops can be performed in an explicitly gauge-invariant fashion using the background field method.
As is well known, for example from the renormalization of chiral perturbation theory \cite{Gasser} and general relativity \cite{tHooft:1974toh}, the background field method allows one to explicitly preserve the underlying symmetry. Renormalization then involves a local Lagrangian with the correct symmetry. As long as the Lagrangian is the most general one consistent with the symmetry, then the divergences can be renormalized into the coefficients of the Lagrangian. The power-counting of the present theory is discussed more explicitly in the Appendix.

This model could be elaborated by adding flavor degrees of freedom. If one gave the scalar a global ``flavor'' quantum number, $i$, such as an $SU(N_f)$ label, leading to the scalar field carrying both the Yang-Mills ``color'' label $a$ and the flavor label $i$,  $\phi^{a,i}$. The matrix $U$ then lives in both color and flavor space
\begin{equation}
U = \exp (i t^a \lambda^i \frac{\phi^{a,i}}{f})
\end{equation}
where $\lambda^i$ are the $SU(N_f)$ matrices. If $N_f$ is too large, asymptotic freedom will be lost. If the $SU(N_f)$ symmetries are dynamically broken, the study of the Goldstone modes could prove interesting.

\section{Using strong interactions to modify the dispersion relation}

Another setting where the free field arguments against quartic propagators may be mitigated is when the interactions modify the dispersion relation at small virtuality. A physically relevant example here concerns gravitational interactions. Consider a strongly interacting SU(N) Yang-Mills theory coupled to the gravitational field, and use the square of the Weyl tensor as the gravitational action. For example, consider the action
\begin{equation}
S= \int d^4x \sqrt{-g} \left[-\frac{1}{4g_N^2}g^{\mu\alpha}g^{\nu\beta} F_{\mu\nu}^a F_{\alpha\beta}^a  - \frac{1}{2\xi^2} C_{\mu\nu\alpha\beta}C^{\mu\nu\alpha\beta} \right]
\end{equation}
where the Weyl tensor is given by
\begin{eqnarray}
C_{\mu\nu\alpha\beta} &=& {R}_{\mu\nu\alpha\beta} -\frac12 \left( {R}_{\mu\alpha} g_{\nu\beta} - {R}_{\nu\alpha} g_{\mu\beta} - {R}_{\mu\beta} g_{\mu\alpha} + {R}_{\nu\beta} g_{\mu\alpha} \right) \ \ \nonumber \\
&+& \frac{{R(g)}}{6}\left(g_{\mu\alpha} g_{\nu\beta} - e_{\nu\alpha} e_{\mu\beta} \right)
\end{eqnarray}
Both terms in this action are conformally invariant. The use of the Weyl tensor squared as the gravitational action is an attractive starting point for gravity because of the conformal symmetry and because it yields a renormalizeable\cite{Stelle:1976gc} and asymptotically free \cite{Fradkin:1981iu} theory. However, when treated on its own, the Weyl action is viewed a sick because it involves higher time derivatives and leads to a quartic propagator.

The combination is asymptotically free in both couplings. While the beta functions will be coupled beyond leading order, at one loop they are decoupled.
\begin{eqnarray}\label{beta}
\beta(g) &=& -\frac{11N}{3\pi^2} g^3 \nonumber \\
\beta(\xi)&=& - \frac{199}{480\pi^2} \xi^3 - \frac{N^2-1}{160\pi^2} \xi^3
 \ \ .
\end{eqnarray}
where the first term in $\beta(\xi)$ comes from graviton loops\cite{Fradkin:1981iu} and the second is the Yang-Mills field. As one runs down from the UV, the two theories will become strongly interacting at different energy scales, which we can refer to as $\Lambda_N$ and $\Lambda_g$ respectively. For the present discussion, let us assume that $\Lambda_N>\Lambda_g$ so that the Yang-Mills theory defines the highest energy scale of the theory, and the gravitational interaction will be relatively weakly interacting when the Yang-Mills field is strongly interacting. However, since the gravitational field is coupled to the Yang-Mills field, the gravitational interaction itself will be modified by the strong interactions. New effective interactions can be induced and by dimensional transmutation they will carry the scale $\Lambda_N$. Since by assumption, it is the largest scale in the theory, it can the be identified with the Planck scale $\Lambda_N\sim M_P$.

In particular, the Yang Mills interaction will modify the graviton dispersion relation, likely converting it to a well behaved spectrum at low energy.
This happens because along with the confinement of the Yang Mills field, corrections to the gravitational interactions are induced. By general
covariance, at low energy these must take the form of a cosmological constant and an Einstein-Hilbert term. By dimensional transmutation\cite{Holdom, Donoghue:2016vck, Donoghue:2016xnh, Adler:1982ri, Zee:1983mj, Smilga:1982se, Elizalde:1994gv, Visser:2002ew, Salvio:2014soa, Einhorn:2014gfa}, these
would carry the scale factor of the Yang Mills theory. That is, we expect that the low energy action, below the confinement scale starts out as
\begin{equation}
S_{low} = \int d^4x \sqrt{-g}~\left[ a\Lambda_N^4 +b \Lambda_N^2 R +...\right]
\end{equation}
where $\Lambda_N$ is the scale of the SU(N) theory and $a,~b$ are constants. The physical spectrum of the interacting theory is then radically different from that suggested by the pure gravitational theory treated perturbatively. In particular, the effective field theory at low energies
starts with the usual ingredients for general relativity, and not the quartic propagators of the original conformal theory. With the induced cosmological constant and Einstein-Hilbert term, the quartic behavior does not hold at low energy.

Lattice studies should be able to simulate non-pertubatively the inducing of the Einstein-Hilbert terms by the Yang-Mills interaction. This is because one does not need to model the full gravitational interaction, but only to treat the gravitational field as an external field while simulating fully the Yang-Mills portion of the theory. A crucial question to address in the simulation is whether the induced Einstein-Hilbert term has the correct sign. In that case, Yang-Mills driven gravity may be a viable ultraviolet QFT completion for gravity.

Holdom and Ren\cite{Holdom} argue that one can replace this external Yang-Mills interaction by the gravitational interaction itself, as curvature squared terms are asymptotically free and hence strongly coupled at $\Lambda_g$. In the present language, this amounts to choosing the Yang-Mills scale below the gravitational scale $\Lambda_N<\Lambda_g$ and identifying $\Lambda_g\sim M_P$. They describe a proposed mechanism for removing the ghost at low energy and generating the usual graviton propagator. This is also deserving of further investigation. The present author has also suggested\cite{Donoghue:2016vck, Donoghue:2016xnh} that the spin connection, treated as an independent gauge field, may be the field whose strong interactions are responsible for inducing the Einstein-Hilbert action.

This situation appears to contain a paradox. The high energy propagators behave as $1/q^4$. At low energy they behave as $q^0$ (the cosmological constant) and  $q^{-2}$, with a small quartic admixture. By the logic of Eq. \ref{quartic} this would imply a massive ghost state. However, when viewed from low energy this ghost state appears to exist at or above the scale of the strong interaction. When one studies the original theory at this high scale, the ghost state does not appear in the spectrum. In the perturbative region, one finds logarithmic corrections to the propagators, but not poles. The quadratic terms in the propagator are only a low energy approximation. This is a common feature within effective field theory. By the uncertainty principle, effects from high energy are local when viewed at low energy, and can be Taylor expanded in the momenta, and hence using operators with extra derivatives. Quantum corrections within the effective field theory requires that higher derivative terms be present in order to absorb the divergences in the perturbative expansion and also to encode finite effects from the full theory. Very often these higher derivative terms bring in higher time derivatives. However, the sickness implied by Ostrogradsky's theorem does not apply. The reasons are understood. When the high energy theory is healthy, the apparent flaws fall outside the reach of the effective field theory and are absent from the full theory\cite{Burgess:2014lwa, Simon:1990ic}.  When treated as an effective field theory at low energy \cite{Donoghue:1994dn}, one need consider only the leading quadratic propagators. By doing so, one can match onto the full theory in an expansion involving the inverse of the heavy scale, without introducing spurious solutions.

To supplement the more complete discussions of \cite{Burgess:2014lwa, Simon:1990ic}, we can also give a simple example of how quartic interactions in the effective theory need not signal a fundamental instability. This can be found in the low energy limit of QED. The vacuum polarization
\begin{equation}
\Pi_{\mu\nu} (q) =\left( q_\mu q_\nu - q^2 \eta_{\mu\nu} \right)\Pi(q)
\end{equation}
has the form
\begin{eqnarray}
\Pi(q)&=& \frac{e^2}{12\pi^2}\left[\frac{1}{\epsilon} +\log 4\pi - \gamma - 6\int_0^1 dx x(1-x) \log\left(\frac{m^2-q^2x(1-x)}{\mu^2}\right)\right] \ \ \nonumber \\
&=& \frac{e^2}{12\pi^2}\left[\frac{1}{\epsilon} +\log 4\pi - \gamma -\log \frac{m^2}{\mu^2} +\frac{q^2}{5 m^2}+...\right]
\end{eqnarray}
The first form here is the one defined for all values of $q^2$, while the second is the expansion at low energies. After renormalization, one
can match the effect of the vacuum polarization to a low energy effective Lagrangian, finding
\begin{equation}
{\cal L}_{eff} = -\frac14 F_{\mu\nu}F^{\mu\nu} -\frac{\alpha}{60\pi m^2}\partial_\lambda F_{\mu\nu}\partial^\lambda F^{\mu\nu}  \ \ .
\end{equation}
Despite having higher-order time derivatives in the last term, QED does not suffer from the Ostrogradsky instability when this is treated as an effective Lagrangian. The denominator structure of the propagator is found to be
\begin{equation}
\frac1{q^2+\frac{\alpha}{15\pi m^2}q^4} = \frac1{q^2} -\frac1{q^2 +\frac{15\pi m^2}{\alpha} }
\end{equation}
which naively has a negative-norm tachyonic pole. However, that indication is of course spurious. The putative pole occurs at a higher energy than that which the effective Lagrangian is valid.   The full theory has a completely different momentum dependence - logarithmic -  in the energy region where the pole was implied, and one never finds such a pole.  When treated as an effective field theory, the effective Lagrangian matches on to the predictions of the full theory, which is well-behaved. However, we should not use the Ostrogradsky Hamiltonian construction at low energy, nor
infer the existence of high energy ghosts from the low energy approximation to the propagator structure.

The gravitational theory based on the Weyl action is in a sense the reverse of the QED example, in that the original action involved the four derivative term and induced effects produce a two derivative action. However, the net effect is the same. The induced action will contain the dominant energy scale, called $\Lambda_N$ above, and hence the putative ghosts in the propagators of the effective field theory will occur at the scale $\Lambda_N$. However, at that scale the effective theory is no longer valid and the perturbative quantum corrections involve logarithms rather than pole behavior.

\section{Comments}

This paper has discussed two examples in which the original theory has four derivatives but whose naive free Hamiltonian is not relevant for the physical spectrum. In one case the original particle does not appear in the spectrum at all, due to confinement. In the other, strong interactions induce a standard Hamiltonian in the low energy limit. Both of the manifestations described are non-perturbative in nature so that they are difficult to address reliably via analytic methods. However, features of such theories have the potential to be simulated on a lattice.

The general analysis of the physical spectrum is through the study of representations of the Lorentz group. The on-shell states satisfy the four-vector condition $p^2=m^2$ which is taken to imply that the fields satisfy a wave equation equivalent to $[\Box +m^2]\phi(x) =0$. However, it is possible that the initial fields defining a theory need not satisfy this relation, only that the on-shell output be equivalent to this. This is a somewhat weaker condition on a successful theory. If some theories with four derivatives are shown to be physically reasonable, then the possibilities for theory construction are significantly expanded.

Both of these theories are highly non-linear. One can expect that this is a common feature for any theory with four derivatives in four dimensions. In such a case the Lagrangian, which is dimension four, gets its overall power from the derivatives, and the fields then carry dimension zero. As such, without some organizing principle there would an infinite tower of operators in the Lagrangian. However, with a symmetry
principle one can reduce the number of available structures.

\section*{Acknowledgements} I would like to thank most particularly J. J. Carrasco and David Kosower for sharing their insights. I have also benefited from discussions with Pierre Vanhove, Bob Holdom and Jing Ren.  This work has been supported in part by the National Science Foundation under grant NSF PHY15-20292.

\section*{Appendix - Power counting in the scalar field renormalization}

In determining which terms in a general Lagrangian are renormalized, one turns to power-counting arguments. These specify the mass dimension of the terms in the Lagrangian which are effected at a given loop order. For theories with quadratic propagators, the power-counting rules are well known. With propagators quartic in the momenta some modifications are necessary, which however are relatively simple. In this Appendix, the power counting for the scalar model of Section 2 is briefly described.

Let us consider a background field expansion of the matrix field
\begin{equation}\label{background}
U = \bar{U} e^{i\tau^a \Delta^a}
\end{equation}
with  $\bar{U}$ being the background field and $\Delta^a$ being the quantum fluctuation. The background field is takes as a solution to the full equations of motion. This retains the full gauge symmetry of the original theory,
\begin{equation}
 \bar{U} \to V\bar{U}V^\dagger ~~~~~~~~     e^{i\tau\cdot \Delta}\to V  e^{i\tau\cdot \Delta}V^\dagger   \ \ .
\end{equation}
When the quantum field is integrated out, the resulting action will be a function of $\bar{U}$ and will retain this symmetry.

The action can be expanded in powers of the background field. The linear term vanishes because of the equations of motion, and we are left with
\begin{equation}
{\cal{L}}(U) = {\cal{L}} (\bar{U}) + \Delta^a {\cal{O}}^{ab}\Delta^b +...
\end{equation}
where in this case ${\cal{O}}^{ab}$ is a fourth order differential operator, which is a function of the backround field and the gauge field ${\cal{O}}^{ab}(\bar{U}, A^a_\mu)$. In complete generality this can be written in the form
\begin{equation}\label{generalfourth}
{\cal{O}}^{ab} = \left[D^2D^2 +A^{\alpha\beta\gamma} D_{\alpha} D_{\beta} D_{\gamma} +B^{\alpha\beta} D_{\alpha} D_{\beta}  + C^\alpha D_\alpha +E\right]^{ab}
\end{equation}
where $D_\alpha$ is the gauge covariant derivative and the functions $A,~B,~C,~ E $ are functions of the background field and their derivatives. In our case, the operator $E$ vanishes, because without any derivative acting on $\Delta$ the background field expansion of Eq. \ref{background} is a symmetry transformation. However we keep $E$ in the operator in order to show that it does not contribute to the divergences. By power-counting $A$ had dimension one, $B$ has dimension two, $C$ has dimension three and $E$ dimension four. Let us refer to the dimension of the operator as $n_i$, such that $n_A=1, ~n_B=2,$ etc. The coefficients $A^{\alpha\beta\gamma}$ and $B^{\alpha\beta} $ are taken to be symmetric in their Lorentz indices. For use below, we define
\begin{equation}
A^\alpha = \eta_{\beta\gamma} A^{\alpha \beta\gamma}~~,~~~~~B= \eta_{\alpha\beta} B^{\alpha \beta}~ \ \ .
\end{equation}

Before present the explicit result, we can see the general structure of the result diagrammatically. If we expand
\begin{equation}\label{expansion}
  {\cal O}^{ab} = \delta^{ab} \Box^2 + V^{ab}
\end{equation}
the one-loop action can be found via
\begin{eqnarray}\label{trlog}
  Tr \log {\cal O} &=&Tr \log (\Box^2 +V) = \left[ Tr \log (\Box^2)  ~+ ~Tr \log (1+ \frac{1}{\Box^2} V ) \right] \nonumber \\
  &=& {\rm const. } + Tr (\frac{1}{\Box^2} V ) + \frac12 Tr (\frac{1}{\Box^2} V \frac{1}{\Box^2} V ) +\frac13 Tr (\frac{1}{\Box^2} V \frac{1}{\Box^2} V\frac{1}{\Box^2} V ) +.... \ \ .
\end{eqnarray}
This generates the tadpole, bubble, triangle, box etc. Feynman diagrams. The massless tadpole always vanishes in dimensional regularization\footnote{For this reason, when we quote the divergent heat kernel coefficient from Ref. \cite{Barvinsky:1985an} below, we drop the result of the tadpole (labeled $\hat{P} $ in that work or $E$ above).  }. The bubble diagram involves the Feynman integral
\begin{equation}\label{bubblefeyn}
  \int \frac{ d^d k}{(2\pi)^d} \frac{k^{4-n_i}(k+q)^{4-n_j}}{k^4 (k+q)^4}
\end{equation}
where $n_i ,~n_j$ are the dimensions of the operators at the two vertices of the bubble diagram and $q$ is the momentum flowing through it. The factors of momenta in the numerator come from the derivatives in the general operator Eq. \ref{generalfourth}. We see that the integral will be
UV divergent if $n_i +n_i \le 4$. The overall power of the integral is set by the momentum $q$, which turns into powers of a derivative when converted into a Lagrangian. From this power-counting, we see that the divergences will correspond to effective operators $A\partial^2 A, ~A \partial B, ~ A C, ~B^2$ with tensor indices and derivatives arranged in various ways. We see that $E$ would not contribute to the divergences, even if it were not vanishing in our model. A similar analysis of the triangle diagram involves 3 vertices, and leads to the conclusion that operators with
$n_i+n_j+n_k \le 4$ are associated with divergences. This includes the factors $A\partial A^2,~ BAA$ only. The box diagram with 4 vertices also leads to a divergence if $n_i+n_j+n_k +n_l\le 4$, which is satisfied only for $A^4$ operators.

The divergences in the functional determinant can be evaluated using the heat kernel expansion. Using
\begin{equation}\label{lnD}
  Tr \log {\cal D} = \int d^dx ~Tr <x| \log {\cal D}|x> = - \int d^dx~ \int_0^\infty \frac{d\tau}{\tau} ~Tr <x|e^{-\tau {\cal D}}|x>
\end{equation}
we can evaluate the action using the heat kernel
\begin{equation}\label{heatkernel}
  K(x,\tau) \equiv <x|e^{-\tau {\cal D}}|x>  \ \ .
\end{equation}
Using the expansion in terms of the Seeley-DeWitt coefficients, $a_i$
\begin{equation}\label{expansion}
  H(x,\tau) = \frac{i}{(4\pi)^{d/2}}\frac{e^{-\tau m^2}}{\tau^{d/2}}\left[a_0(x)  +a_1(x) \tau + a_2(x) \tau^2 +... \right]
\end{equation}
We can identify the divergent part of the functional determinant residing in the $a_2$ coefficient
\begin{equation}\label{divergent}
  Tr <x| \log {\cal D}|x>|_{div} = \frac{i}{(4\pi)^{d/2}} \Gamma(2-\frac{d}{2}) ~Tr~ a_2 (x)
\end{equation}

The Seeley-DeWitt coefficients for a quadratic differential operator are simple and well known. For the quartic operator of Eq. \ref{generalfourth}, the coefficients are more complicated and have been given by Barvinsky and Vilkovisky in Ref. \cite{Barvinsky:1985an}.
In flat space the $a_2$ coefficient, which controls the divergences, has the form
\begin{eqnarray}
a_2 &=& \frac{1}{6} F_{\mu\nu}F^{\mu\nu} -\frac18 F_{\mu\nu}[D^\mu,A^\nu]+\frac{9}{80}F_{\mu\nu}A^{\mu\alpha\beta}A^\nu_{~\alpha\beta} +\frac{9}{160}F_{\mu\nu}A^\mu A^\nu +\frac18 C_\mu A^\mu + \frac{1}{24} B_{\mu\nu}B^{\mu\nu} +\frac{1}{48} B^2  \nonumber \\ &-& \frac{1}{16} B D_\mu A^\mu +\frac18 B^{\mu\nu} D_\mu A_\nu -\frac18 B^{\mu\nu}D^\alpha A_{\mu\nu\alpha} +\frac{9}{80}A^{\mu\nu\alpha}D_\mu D_\nu A_\alpha -\frac{3}{80} A^\mu D_\mu D_\nu A^\nu \nonumber -\frac{3}{160} A^\mu D^2 A_\mu \\ &-&\frac{3}{40}A^{\mu\nu\alpha}D_\mu D_\beta   A^\beta_{~\nu\alpha}-\frac{1}{80}A^{\mu\nu\alpha}D^2 A_{\mu\nu\alpha} -\frac{1}{640}B(2A^{\mu\nu\alpha}A_{\mu\nu\alpha}+3A^\mu A_\mu)
\nonumber \\ &-&\frac{3}{320}B^{\mu\nu}(2 A_{\mu\alpha\beta}A_\nu^{~\alpha\beta}+ A_\mu A_\nu + A_{\mu\nu\alpha}A^\alpha + A^\alpha A_{\mu\nu\alpha})  -\frac{1}{640}A^\beta D_\beta ( 2  A^{\mu\alpha\beta} A_{\mu\alpha\beta}+ 3A^\mu A_\mu ) \nonumber \\
&-&\frac{1}{640}A^\beta  [ 2(D_\beta  A^{\mu\nu\alpha}) A_{\mu\nu\alpha}+ 3 (D_\beta A^\mu) A_\mu]- \frac{3}{160}
A^{\mu\nu\alpha} D_\mu(A_\nu A_\alpha +2A_{\nu\beta\gamma} A_\alpha^{~\beta\gamma}+A_{\nu\alpha\beta} A^\beta + A^\beta A_{\nu\alpha\beta}) \nonumber \\
 &+& \frac{3}{320}A^{\mu\nu\alpha} (A_\nu D_\mu A_\alpha +2A_{\nu\beta\gamma}D_\mu A_\alpha^{~\beta\gamma}+A_{\nu\alpha\beta}D_\mu A^\beta + A^\beta D_\mu A_{\nu\alpha\beta}) \nonumber \\
&+& \frac{1}{960 \times 32 \times 42} A^{\mu\nu\alpha} A^{\beta \gamma \delta} A^{\epsilon\sigma\rho} A^{\lambda\omega\eta} g_{\mu\nu\alpha\beta\gamma\epsilon\sigma\rho\lambda\omega\eta}
\end{eqnarray}
where $g_{\mu\nu\alpha\beta\gamma\epsilon\sigma\rho\lambda\omega\eta}$ is the completely symmetric tensor
\begin{equation}
g_{\mu\nu\alpha\beta\gamma\delta\epsilon\sigma\rho\lambda\omega\eta} = g_{\mu\nu}g_{\alpha\beta}g_{\gamma\delta}g_{\epsilon\sigma}g_{\rho\lambda}g_{\omega\eta} + g_{\mu\alpha}g_{\nu\beta}g_{\gamma\delta}g_{\epsilon\sigma}g_{\rho\lambda}g_{\omega\eta} + ~{\rm perms.}
\end{equation}
with a total of $11!! = 10,395$ terms.

It is easy to verify that all of the terms in this result are of dimension four. The result can be decomposed into terms in the most general dimension four Lagrangian, which is the original Lagrangian of the theory. While the computation is an algebraic nightmare, the background field method plus power counting verifies that the one-loop divergences can be absorbed into the renormalization of the parameters of the most general dimension-four Lagrangian.


\begin{thebibliography}{99}

\bibitem{Woodard:2015zca}
  R.~P.~Woodard,
  ``Ostrogradsky's theorem on Hamiltonian instability,''
  Scholarpedia {\bf 10}, no. 8, 32243 (2015)
  doi:10.4249/scholarpedia.32243
  [arXiv:1506.02210 [hep-th]].









\bibitem{Tomboulis:1983sw}
  E.~T.~Tomboulis,
  ``Unitarity in Higher Derivative Quantum Gravity,''
  Phys.\ Rev.\ Lett.\  {\bf 52}, 1173 (1984).
  doi:10.1103/PhysRevLett.52.1173 \\
  I.~Antoniadis and E.~T.~Tomboulis,
  ``Gauge Invariance and Unitarity in Higher Derivative Quantum Gravity,''
  Phys.\ Rev.\ D {\bf 33}, 2756 (1986).
  doi:10.1103/PhysRevD.33.2756

\bibitem{Smilga}
  A.~V.~Smilga,
  ``Benign versus malicious ghosts in higher-derivative theories,''
  Nucl.\ Phys.\ B {\bf 706}, 598 (2005)
  [hep-th/0407231].\\
  A.~V.~Smilga,
  ``Quantum gravity as Escher's dragon,''
  Phys.\ Atom.\ Nucl.\  {\bf 66}, 2092 (2003)
  [Yad.\ Fiz.\  {\bf 66}, 2141 (2003)]
  [hep-th/0212033]. \\
 A.~V.~Smilga,
  ``Ghost-free higher-derivative theory,''
  Phys.\ Lett.\ B {\bf 632}, 433 (2006)
  [hep-th/0503213].

\bibitem{Bender:2007wu}
  C.~M.~Bender and P.~D.~Mannheim,
  ``No-ghost theorem for the fourth-order derivative Pais-Uhlenbeck oscillator model,''
  Phys.\ Rev.\ Lett.\  {\bf 100}, 110402 (2008)
  doi:10.1103/PhysRevLett.100.110402
  [arXiv:0706.0207 [hep-th]]. \\
 C.~M.~Bender and P.~D.~Mannheim,
  ``Giving up the ghost,''
  J.\ Phys.\ A {\bf 41}, 304018 (2008)
  doi:10.1088/1751-8113/41/30/304018
  [arXiv:0807.2607 [hep-th]]. \\
    C.~M.~Bender and P.~D.~Mannheim,
  ``Exactly solvable PT-symmetric Hamiltonian having no Hermitian counterpart,''
  Phys.\ Rev.\ D {\bf 78}, 025022 (2008)
  doi:10.1103/PhysRevD.78.025022
  [arXiv:0804.4190 [hep-th]].


\bibitem{Salvio:2015gsi}
  A.~Salvio and A.~Strumia,
  ``Quantum mechanics of 4-derivative theories,''
  Eur.\ Phys.\ J.\ C {\bf 76}, no. 4, 227 (2016)
  doi:10.1140/epjc/s10052-016-4079-8
  [arXiv:1512.01237 [hep-th]].
\bibitem{Mottola:2016mpl}
  E.~Mottola,
  ``Scalar Gravitational Waves in the Effective Theory of Gravity,''
  arXiv:1606.09220 [gr-qc].

\bibitem{Holdom}
 B.~Holdom and J.~Ren,
  ``QCD analogy for quantum gravity,''
  Phys.\ Rev.\ D {\bf 93}, no. 12, 124030 (2016)
  doi:10.1103/PhysRevD.93.124030
  [arXiv:1512.05305 [hep-th]]. \\
  B.~Holdom and J.~Ren,
  ``Quadratic gravity: from weak to strong,''
  Int.\ J.\ Mod.\ Phys.\ D {\bf 25}, no. 12, 1643004 (2016)
  doi:10.1142/S0218271816430045
  [arXiv:1605.05006 [hep-th]].

\bibitem{Donoghue:2016vck}
  J.~F.~Donoghue,
  ``Is the spin connection confined or condensed?,''
  arXiv:1609.03523 [hep-th].

\bibitem{Donoghue:2016xnh}
  J.~F.~Donoghue,
  ``A conformal model of gravitons,''
  arXiv:1609.03524 [hep-th].



\bibitem{Gasser}
  J.~Gasser and H.~Leutwyler,
  ``Chiral Perturbation Theory: Expansions in the Mass of the Strange Quark,''
  Nucl.\ Phys.\ B {\bf 250}, 465 (1985).
  doi:10.1016/0550-3213(85)90492-4

\bibitem{tHooft:1974toh}
  G.~'t Hooft and M.~J.~G.~Veltman,
  ``One loop divergencies in the theory of gravitation,''
  Ann.\ Inst.\ H.\ Poincare Phys.\ Theor.\ A {\bf 20}, 69 (1974).






\bibitem{Stelle:1976gc}
  K.~S.~Stelle,
  ``Renormalization of Higher Derivative Quantum Gravity,''
  Phys.\ Rev.\ D {\bf 16}, 953 (1977).
  doi:10.1103/PhysRevD.16.953
\bibitem{Barth:1983hb}
  N.~H.~Barth and S.~M.~Christensen,
  ``Quantizing Fourth Order Gravity Theories. 1. The Functional Integral,''
  Phys.\ Rev.\ D {\bf 28}, 1876 (1983).
  doi:10.1103/PhysRevD.28.1876

\bibitem{Fradkin:1981iu}
  E.~S.~Fradkin and A.~A.~Tseytlin,
  ``Renormalizable asymptotically free quantum theory of gravity,''
  Nucl.\ Phys.\ B {\bf 201}, 469 (1982).
  doi:10.1016/0550-3213(82)90444-8
\bibitem{Avramidi:1985ki}
  I.~G.~Avramidi and A.~O.~Barvinsky,
  ``Asymptotic Freedom In Higher Derivative Quantum Gravity,''
  Phys.\ Lett.\ B {\bf 159}, 269 (1985).
  doi:10.1016/0370-2693(85)90248-5

\bibitem{Adler:1982ri}
  S.~L.~Adler,
  ``Einstein Gravity as a Symmetry Breaking Effect in Quantum Field Theory,''
  Rev.\ Mod.\ Phys.\  {\bf 54}, 729 (1982)
  Erratum: [Rev.\ Mod.\ Phys.\  {\bf 55}, 837 (1983)].
  doi:10.1103/RevModPhys.54.729

\bibitem{Zee:1983mj}
  A.~Zee,
  ``Einstein Gravity Emerging From Quantum Weyl Gravity,''
  Annals Phys.\  {\bf 151}, 431 (1983).
  doi:10.1016/0003-4916(83)90286-5
\bibitem{Smilga:1982se}
  A.~V.~Smilga,
  ``Spontaneous generation of the Newton constant in the renormalizable gravity theory,''
  IN *ZVENIGOROD 1982, PROCEEDINGS, GROUP THEORETICAL METHODS IN PHYSICS, VOL. 2* 73-77.
  [arXiv:1406.5613 [hep-th]].

\bibitem{Elizalde:1994gv}
  E.~Elizalde, S.~D.~Odintsov and A.~Romeo,
  ``Improved effective potential in curved space-time and quantum matter, higher derivative gravity theory,''
  Phys.\ Rev.\ D {\bf 51}, 1680 (1995)
  doi:10.1103/PhysRevD.51.1680
  [hep-th/9410113].


\bibitem{Visser:2002ew}
  M.~Visser,
  ``Sakharov's induced gravity: A Modern perspective,''
  Mod.\ Phys.\ Lett.\ A {\bf 17}, 977 (2002)
  doi:10.1142/S0217732302006886
  [gr-qc/0204062].



\bibitem{Salvio:2014soa}
  A.~Salvio and A.~Strumia,
  ``Agravity,''
  JHEP {\bf 1406}, 080 (2014)
  doi:10.1007/JHEP06(2014)080
  [arXiv:1403.4226 [hep-ph]].





\bibitem{Einhorn:2014gfa}
  M.~B.~Einhorn and D.~R.~T.~Jones,
  ``Naturalness and Dimensional Transmutation in Classically Scale-Invariant Gravity,''
  JHEP {\bf 1503}, 047 (2015)
  doi:10.1007/JHEP03(2015)047
  [arXiv:1410.8513 [hep-th]].\\
 T.~Jones and M.~Einhorn,
  ``Quantum Gravity and Dimensional Transmutation,''
  PoS PLANCK {\bf 2015}, 061 (2015).














\bibitem{Burgess:2014lwa}
  C.~P.~Burgess and M.~Williams,
  ``Who You Gonna Call? Runaway Ghosts, Higher Derivatives and Time-Dependence in EFTs,''
  JHEP {\bf 1408}, 074 (2014)
  doi:10.1007/JHEP08(2014)074
  [arXiv:1404.2236 [gr-qc]].\\
  C.~P.~Burgess,
  ``Introduction to Effective Field Theory,''
  Ann.\ Rev.\ Nucl.\ Part.\ Sci.\  {\bf 57}, 329 (2007)
  doi:10.1146/annurev.nucl.56.080805.140508
  [hep-th/0701053].

\bibitem{Simon:1990ic}
  J.~Z.~Simon,
  ``Higher Derivative Lagrangians, Nonlocality, Problems and Solutions,''
  Phys.\ Rev.\ D {\bf 41}, 3720 (1990).
  doi:10.1103/PhysRevD.41.3720


\bibitem{Donoghue:1994dn}
  J.~F.~Donoghue,
  ``General Relativity As An Effective Field Theory: The Leading Quantum
  Corrections,''
  Phys.\ Rev.\  D {\bf 50}, 3874 (1994)
  [arXiv:gr-qc/9405057].

\bibitem{Barvinsky:1985an}
  A.~O.~Barvinsky and G.~A.~Vilkovisky,
  ``The Generalized Schwinger-Dewitt Technique in Gauge Theories and Quantum Gravity,''
  Phys.\ Rept.\  {\bf 119}, 1 (1985).
  doi:10.1016/0370-1573(85)90148-6

\end{thebibliography}
\end{document}